\documentclass[12pt]{article}

\usepackage[utf8]{inputenc}
\usepackage[T1]{fontenc}
\usepackage[english]{babel}
\usepackage[a4paper, margin=1in]{geometry}
\usepackage{setspace}
\usepackage{amsmath, amssymb, amsfonts, amsthm, mathrsfs, mathtools}
\usepackage{bm}
\usepackage{slashed}
\usepackage{physics}
\usepackage{tensor}
\usepackage{textgreek}
\usepackage{braket}
\usepackage{cancel}
\usepackage{esint}
\usepackage{bbold}
\usepackage{dsfont}
\usepackage{upgreek}
\usepackage{stmaryrd}
\usepackage{enumitem}
\usepackage{graphicx}
\usepackage{tikz}
\usepackage{pgfplots}
\usepackage{tikz-cd}
\usepackage{tikz-3dplot}
\usepackage{xcolor}
\usepackage{float}
\usepackage{caption}
\usepackage{subcaption}
\usepackage{hyperref}
\usepackage{cleveref}
\usepackage{fancyhdr}
\usepackage{appendix}
\usepackage{titlesec}
\usepackage{authblk}
\usepackage{cite}
\usepackage{comment}
\usepackage{multicol}
\usepackage{lipsum}
\usepackage{etoolbox}
\usepackage{array}
\usepackage{longtable}
\usepackage{booktabs}
\usepackage{makecell}
\usepackage{colortbl}
\usepackage{multirow}
\usepackage{pdfpages}
\usepackage{mathptmx} 

\pgfplotsset{compat=newest}

\hypersetup{
  colorlinks=true,
  linkcolor=blue,
  citecolor=blue,
  filecolor=blue,
  urlcolor=blue
}

\title{\textbf{\huge On Immutable Memory Systems for Artificial Agents:\\ A Blockchain-Indexed Automata-Theoretic Framework Using ECDH-Keyed Merkle Chains}}

\author{\Large Dr Craig S. Wright\\
\small University of Exeter Business School\\
\small Exeter, United Kingdom\\
\small \texttt{cw881@exeter.ac.uk}}

\date{June 16, 2025}

\begin{document}

\maketitle

\begin{abstract}
This paper presents a formalised architecture for synthetic agents designed to retain immutable memory, verifiable reasoning, and constrained epistemic growth. Traditional AI systems rely on mutable, opaque statistical models prone to epistemic drift and historical revisionism. In contrast, we introduce the concept of the \emph{Merkle Automaton}—a cryptographically anchored, deterministic computational framework that integrates formal automata theory with blockchain-based commitments. Each agent transition, memory fragment, and reasoning step is committed within a Merkle structure rooted on-chain, rendering it non-repudiable and auditably permanent.

To ensure selective access and confidentiality, we derive symmetric encryption keys from ECDH exchanges contextualised by hierarchical privilege lattices. This enforces cryptographic access control over append-only DAG-structured knowledge graphs. Reasoning is constrained by formal logic systems and verified through deterministic traversal of policy-encoded structures. Updates are non-destructive and historied, preserving epistemic lineage without catastrophic forgetting. Zero-knowledge proofs facilitate verifiable, privacy-preserving inclusion attestations. Collectively, this architecture reframes memory not as a cache but as a ledger—one whose contents are enforced by protocol, bound by cryptography, and constrained by formal logic.

The result is not an intelligent agent that mimics thought, but an epistemic entity whose outputs are provably derived, temporally anchored, and impervious to post hoc revision. This design lays foundational groundwork for legal, economic, and high-assurance computational systems that require provable memory, unforgeable provenance, and structural truth.
\end{abstract}


\newpage


\section{Introduction}

In modern computing, the integrity of stateful agents depends not solely upon performance metrics or heuristic accuracy but upon their capacity to persist and justify memory over time. The architectural challenge lies in preventing epistemic drift—wherein synthetic agents subtly mutate their stored beliefs through stochastic updates, model degradation, or retraining processes that fail to preserve past commitments. While traditional neural systems capture statistical approximations of reality, they are unable to preserve the chain of reasoning that led to any particular output. Consequently, trust becomes probabilistic, grounded in correlations rather than in proof. Such architectures render agents opaque, unverifiable, and fundamentally unreliable in contexts requiring accountability.

This paper introduces a mathematically rigorous, cryptographically anchored agent design based on finite-state machines extended with memory permanence, zero-knowledge auditability, and formal policy derivation. We propose a novel formalism—termed the Merkle Automaton—that extends classical automata theory with blockchain-based commitment schemes, enabling each transition, belief, and output to be irreversibly recorded, cryptographically sealed, and externally verified. In this paradigm, memory is not ephemeral nor mutable but instead becomes a substrate of law, governed by access privileges, formal ontologies, and deductive constraints.

Rooted in formal language theory and inspired by secure ledger structures, our framework incorporates ECDH-based symmetric encryption for controlled access, layered DAG topologies for epistemic evolution, and zero-knowledge proofs for selective disclosure. We argue that the combination of deterministic automata and cryptographic anchoring allows agents to serve as verifiable epistemic actors whose outputs can be traced, justified, and audited. Furthermore, we show how reasoning becomes an act of DAG traversal, constrained by append-only knowledge commitments and subjected to formal verification rules.

This architecture yields not intelligence in the anthropomorphic sense, but structural truth: agents that cannot hallucinate, cannot fabricate, and cannot revise history. Instead, they operate under deductive legality. Their memory is not merely accessed but proven. This, we contend, is the foundational precondition for artificial agents operating in regulated, adversarial, or high-assurance domains.

\section{The Problem of Epistemic Drift in Synthetic Agents}

Synthetic agents that rely on stochastic embedding-based systems exhibit epistemic drift—defined as the progressive, untraceable deviation between output assertions and any fixed epistemic foundation. This arises due to the mutable nature of model weights, the continuous updating of high-dimensional vector spaces, and the lack of cryptographically verifiable memory. Such systems generate outputs based on probabilistic proximity in latent space rather than on reproducible state-based reasoning.

Let \( f_\theta: \mathcal{X} \rightarrow \mathcal{Y} \) denote a neural approximation function parameterised by \( \theta \in \mathbb{R}^n \). For any input \( x \in \mathcal{X} \), the output \( y = f_\theta(x) \) is a function of the current state of the weights \( \theta \), which evolve during training according to gradient descent or other heuristic algorithms. However, \( \theta \) lacks historical grounding. Once updated, the previous epistemic trajectory becomes irrecoverable unless manually versioned, which is not enforced by any formal constraint in the architecture.

This leads to the phenomenon of epistemic non-repeatability:
\[
f_{\theta_t}(x) \neq f_{\theta_{t+\Delta t}}(x) \quad \text{even when } x \text{ is held constant},
\]
where \( \theta_t \) and \( \theta_{t+\Delta t} \) are successive model states. This violates the basic condition of epistemic traceability, wherein identical queries should yield consistent justifications traceable to past states.

From a formal logic standpoint, traditional epistemic logic (cf. Hintikka, 1962) assumes modal operators \( K_i \varphi \), read as "agent \( i \) knows \( \varphi \)," to be closed under logical consequence. However, in neural agents, no such closure is guaranteed. Further, due to lack of memory anchoring, no Kripke-style accessibility relation \( R \subseteq W \times W \) (where \( W \) is the set of possible worlds) can be meaningfully constructed, as the agent’s world model is mutable and opaque.

In contrast, the mathematical theory of information proposed by Shannon enforces structural invariance through entropy-preserving channels \cite{shannon1948}. For agents to maintain epistemic consistency, they must operate as structured information systems where knowledge is derivable, reproducible, and commit-referenced. Drift represents a violation of these conditions.

Moreover, dynamic embeddings induce topological instability. Consider a high-dimensional manifold \( \mathcal{M} \subseteq \mathbb{R}^n \) encoding concept representations. Continuous training shifts \( \mathcal{M} \), thereby re-mapping the geodesics between concepts. Let \( d_t(p, q) \) denote the geodesic distance between concepts \( p \) and \( q \) at time \( t \). Then:
\[
d_t(p, q) \neq d_{t+\Delta t}(p, q),
\]
implying that semantic relationships are not conserved. This leads to epistemological incoherence: the agent reinterprets prior inputs using altered topologies without version control.

Philosophically, this contravenes Popperian falsifiability. A falsifiable system must retain the epistemic structures against which hypotheses are tested. Agents suffering from epistemic drift effectively erase or morph these structures during training. Hence, they cannot engage in genuine hypothesis testing because the referential substrate shifts over time.

Recent empirical evaluations demonstrate this directly. Large-scale language models fail to maintain fact consistency even under minimal perturbation. In zero-shot settings, factual recall degrades over successive iterations—demonstrating epistemic instability not attributable to adversarial noise, but to the architecture's incapacity for immutable knowledge anchoring \cite{li2023ephemeral}.

A proper formal agent must instead preserve a fixed mapping:
\[
\text{Input} \rightarrow \text{Transition} \rightarrow \text{Justified Output},
\]
where each transition is bound to a cryptographic proof. Epistemic drift is mathematically eliminated when transitions are recorded immutably, such that the entire reasoning path is recoverable and verifiable.

Thus, any architecture aspiring toward cognitive reliability must reject gradient-trained stochastic embeddings as the core epistemic mechanism. Instead, systems must be designed atop deterministic automata, ledger-anchored transitions, and cryptographically enforced referential permanence.

\section{Foundations in Automata and Language Recognition}

To construct epistemically constrained artificial agents, one must begin not with data but with the axiomatic formulation of computation itself. The automaton, a mathematical abstraction first formalised in the mid-20th century, offers the only viable scaffold for grounding state transitions in logic rather than in statistical hallucination.

Formally, a deterministic finite automaton (DFA) is a quintuple:
\[
A = (Q, \Sigma, \delta, q_0, F),
\]
where:
\begin{itemize}
    \item \( Q \) is a finite set of states,
    \item \( \Sigma \) is a finite input alphabet,
    \item \( \delta: Q \times \Sigma \rightarrow Q \) is the deterministic transition function,
    \item \( q_0 \in Q \) is the start state,
    \item \( F \subseteq Q \) is the set of accepting states.
\end{itemize}

The extended transition function \( \delta^* \) over strings is defined recursively by:
\[
\delta^*(q, \varepsilon) = q, \quad \delta^*(q, xa) = \delta(\delta^*(q, x), a) \text{ for } x \in \Sigma^*, a \in \Sigma.
\]
The language recognised by \( A \) is:
\[
L(A) = \{ w \in \Sigma^* \mid \delta^*(q_0, w) \in F \}.
\]

However, such machines are inadequate for epistemic traceability. A DFA forgets its past by construction. Its states do not encode history. They merely collapse all computational antecedents into a single node. No mechanism exists to retrieve the input sequence that produced the current state.

Hence, we propose an augmentation:
\[
A' = (Q, \Sigma, \delta, q_0, F, T),
\]
where \( T \subseteq Q \times \Sigma \times Q \times \mathbb{T} \times \mathbb{H} \) is a transition ledger with time-indexed, hashed transitions. Each tuple \( (q_i, \sigma, q_j, t_i, h_i) \in T \) encodes the transition \( q_i \xrightarrow{\sigma} q_j \) at time \( t_i \) with hash \( h_i = H(q_i \parallel \sigma \parallel q_j \parallel t_i) \).

This modification induces a non-forgetful automaton. The state of the machine is no longer atomic; it is the end of a chain. Let \( \gamma_n = (q_0, \sigma_1, q_1, \ldots, \sigma_n, q_n) \) be the trace of execution. Then:
\[
\forall i \in \{1, \ldots, n\}, \exists t_i, h_i \text{ such that } H(q_{i-1} \parallel \sigma_i \parallel q_i \parallel t_i) = h_i,
\]
and \( (q_{i-1}, \sigma_i, q_i, t_i, h_i) \in T \).

Define the Merkle path \( \mathcal{P}_n \) over \( \{ h_1, \ldots, h_n \} \) and let \( R_n = \text{Root}(\mathcal{P}_n) \). When \( R_n \) is anchored in an externally verifiable blockchain block \( B_n \), then the execution trace becomes immutable and externally attestable.

Thus, the language accepted by this system becomes not merely the set of strings accepted by a transition path to an accepting state, but the set of all verifiable, time-anchored transition sequences leading to such states. We define the verifiable language:
\[
L_{\text{ver}}(A') = \left\{ w \in \Sigma^* \;\middle|\; \exists \gamma_n, \mathcal{P}_n \text{ s.t. } \delta^*(q_0, w) \in F \land \text{MerkleVerify}(\gamma_n, \mathcal{P}_n) = \texttt{true} \right\}.
\]

This transformation shifts the automaton from a black-box recogniser to a white-box ledger of epistemic transitions. It encodes not just acceptance but justification—transforming language recognition from mere pattern matching to logically and cryptographically certified cognition.

\section{The Merkle Automaton: Structural Anchoring of State}

The deterministic automaton, traditionally framed as a quintuple \((Q, \Sigma, \delta, q_0, F)\), where \(Q\) is the set of states, \(\Sigma\) the input alphabet, \(\delta: Q \times \Sigma \rightarrow Q\) the transition function, \(q_0\) the start state, and \(F \subseteq Q\) the accepting states, offers a formal foundation for computation. However, the limitations of this construct become apparent when provenance and tamper-evidence are necessary properties for transition validation. In this section, we formalise the Merkle Automaton, an automaton whose transition structure is rooted not merely in computation, but in verifiable state commitment.

Let each transition \(\delta(q_i, \sigma_j) = q_k\) emit an output \(o_n\). The tuple \((q_i, \sigma_j, q_k, o_n)\) defines a transition-event. We construct a Merkle tree \(M_n\) at each \(n\)-th transition point, where the leaves are all valid output events at that point in time. The root hash \(\text{Root}(M_n)\) is inserted into a public blockchain transaction, denoted \(\mathcal{B}_n\), such that:

\[
\mathcal{B}_n = \text{TX}_{n}(\text{Root}(M_n))
\]

This commitment enables the derivation of a proof-of-inclusion path \(\pi(o_n)\) for any given output, permitting post hoc validation of the automaton’s behaviour. The automaton thus transitions from an abstract state machine to a cryptographically verifiable execution trace machine.

Define the extended Merkle Automaton as the septuple:

\[
\mathcal{A}_\mathcal{M} = (Q, \Sigma, \delta, q_0, F, O, \mathcal{M})
\]

where:

\begin{itemize}
  \item \(Q\): Finite set of states
  \item \(\Sigma\): Input alphabet
  \item \(\delta\): Transition function with output
  \item \(q_0\): Initial state
  \item \(F\): Accepting states
  \item \(O\): Output set where \(o_n \in O\)
  \item \(\mathcal{M}\): Sequence \((M_1, M_2, \dots, M_n)\) of Merkle trees rooted on a blockchain ledger
\end{itemize}

Each \(M_n\) is constructed as:

\[
M_n = \text{MerkleTree}(\{ H(o^n_1), H(o^n_2), \dots, H(o^n_k) \})
\]

with \(H\) being a collision-resistant cryptographic hash function, e.g., SHA-256. The Merkle root \(R_n = \text{Root}(M_n)\) is committed in a blockchain transaction \(\mathcal{B}_n\), rendering the automaton's outputs non-repudiable.

A formal output trace \(\tau = (o_1, o_2, \dots, o_n)\) is considered valid if and only if:

\[
\forall o_i \in \tau,\ \exists\ \pi(o_i): \text{Verify}(\pi(o_i), R_i) = \texttt{True}
\]

where \(\pi(o_i)\) is the Merkle inclusion proof for \(o_i\) in \(M_i\), and \(\text{Verify}\) is the standard Merkle proof verification function.

This construction aligns computational output with formal properties of distributed integrity. It prevents speculative hallucination by requiring that any generated token or output be backed by a verifiable execution trace. If an agent produces a token \(t\) such that:

\[
\neg \exists\ \pi(t): \text{Verify}(\pi(t), R_t) = \texttt{True}
\]

then \(t\) is by construction unverifiable, and may be discarded under a policy of epistemic minimalism.

Let us define the language recognised by the Merkle Automaton as:

\[
\mathcal{L}_\mathcal{M} = \{ w \in \Sigma^* \mid \exists\ q_f \in F,\ \delta^*(q_0, w) = q_f\ \land\ \forall o_i \in \tau(w),\ \text{Verify}(\pi(o_i), R_i) = \texttt{True} \}
\]

where \(\delta^*\) is the extended transition function, and \(\tau(w)\) the output trace generated by \(w\).

In effect, \(\mathcal{L}_\mathcal{M}\) is not merely a recognisable language in the classical Chomskyan hierarchy, but a verifiable language — every accepted string is backed by cryptographic evidence of its generative path.

This approach renders the automaton’s operational semantics cryptographically accountable. Transitions become historically fixed, and outputs immutable unless forked at the blockchain layer, which introduces a measurable cost and systemic resistance.

The Merkle Automaton therefore transcends the classical model by incorporating principles of cryptographic commitment and distributed consensus, providing a foundation for agent-based reasoning systems that do not merely compute, but attest.

\section{The Information Substrate—Blockchain as Oracle of Record}

Traditional memory models in machine intelligence rely upon mutable internal representations—vectors in high-dimensional space adjusted by backpropagation. Such structures, while enabling generalisation, lack any form of epistemic accountability. In contrast, we define a formal substrate where each knowledge fragment is encoded, committed, and verifiable, with its existence and origin independently attestable. The blockchain here is not merely a distributed ledger—it is a cryptographic oracle of record, binding assertion to history.

Let each knowledge unit be a tuple:

\[
K_i = \left( d_i, p_i, H(d_i \parallel p_i), E_{k_i}(d_i) \right)
\]

where:

\begin{itemize}
  \item \(d_i\) is the data payload (a sentence, symbol, or propositional fragment),
  \item \(p_i\) is its provenance (source identifier, timestamp, originator signature),
  \item \(H(d_i \parallel p_i)\) is a cryptographic hash computed over the concatenation of data and provenance,
  \item \(E_{k_i}(d_i)\) is the encrypted form of \(d_i\) under key \(k_i\), derived via elliptic curve cryptographic protocols (cf. ECDH, see §5).
\end{itemize}

Each \(H(d_i \parallel p_i)\) is committed into a Merkle tree \(M_n\), where \(n\) indexes the moment of commitment. The Merkle root \(\text{Root}(M_n)\) is embedded within a transaction \(T_n\) in a blockchain block \(B_n\). Hence, the ledgered inclusion:

\[
B_n \supset T_n(\text{Root}(M_n)) \ni H(d_i \parallel p_i)
\]

binds the knowledge fragment to a cryptographically sealed historical record. The proof-of-inclusion \(\pi(H(d_i \parallel p_i))\) provides verifiability of both existence and order in the temporal structure of the ledger.

From this, we define a function:

\[
\text{VerifyFragment}(K_i, B_n) \Rightarrow \begin{cases}
\texttt{True} & \text{if } \pi(H(d_i \parallel p_i)) \subset \text{MerklePath}(M_n) \land \text{Root}(M_n) \in T_n \in B_n \\
\texttt{False} & \text{otherwise}
\end{cases}
\]

Such a framework transforms memory from an epistemically soft function to a formal process of archival commitment. The concept of forgetting becomes computable: only by invalidating blockchain consensus or removing a committed transaction—both infeasible under adversarial assumptions—can knowledge be erased.

Moreover, the data provenance \(p_i\) must itself be structured. Let us define \(p_i = (s_i, t_i, \sigma_i)\), where:

\begin{itemize}
  \item \(s_i\) is the source document identifier (e.g., DOI, URL hash),
  \item \(t_i\) is the timestamp of observation or recording,
  \item \(\sigma_i\) is the digital signature of the originator (human or agent).
\end{itemize}

The combination \(d_i \parallel p_i\) is not merely a payload—it is a claim. The hash \(H(d_i \parallel p_i)\) then acts as a commitment to that claim, with verifiable provenance and fixed temporal index.

This system allows the model to query its own record as an oracle. A query \(q\) becomes a Merkle-inclusion test, and a response is constrained by verifiable antecedents. No fragment may be hallucinated. The model must derive all inferences from provably committed facts. Let \(R_q\) be the response to query \(q\). Then:

\[
\forall \phi \in R_q,\ \exists\ K_i: \text{VerifyFragment}(K_i, B_n) = \texttt{True} \land \phi \models d_i
\]

This enforces that every statement \(\phi\) output in response to a query must be derivable from committed fragments \(d_i\), with valid Merkle inclusion in the ledger. Memory becomes not a mutable function of training, but a verifiable corpus of ledgered claims.

In doing so, we instantiate a model where the blockchain is not a side-channel of accountability—it is the very substrate of memory. The language of the system is bounded by what can be proven to exist, by what has been sealed in time. The ledger is not metadata; it is the epistemological ground truth.

\section{Symmetric Cryptography from ECDH-Derived Shared Secrets}

Within any epistemically rigid system, data confidentiality must coexist with verifiability. While Merkle inclusion ensures the integrity and anchoring of knowledge fragments, the actual contents may remain confidential to authorised actors. This necessitates a dual system: public auditability and private access. We achieve this through the application of symmetric cryptography, with keys derived from elliptic curve Diffie–Hellman (ECDH) exchanges between long-term asymmetric keypairs.

Let each entity \( U \) (user) and \( A \) (agent) possess a static elliptic curve keypair:

\[
(sk_U, pk_U),\quad (sk_A, pk_A)
\]

The shared secret \( K_{U,A} \) is derived via:

\[
K_{U,A} = \text{ECDH}(sk_U, pk_A) = \text{ECDH}(sk_A, pk_U)
\]

Under the hardness of the Elliptic Curve Diffie–Hellman problem over a secure curve (e.g., Curve25519), this ensures that no third party can derive \( K_{U,A} \) without knowledge of one party's private key.

To avoid direct usage of the raw shared secret, we derive symmetric keys using a context-specific key derivation function (HKDF):

\[
K_{\mathrm{sym},i}^{l} = \text{HKDF}(K_{U,A} \parallel \text{context}_i \parallel l)
\]

where:

\begin{itemize}
  \item \( \text{context}_i \) includes fragment-specific metadata such as timestamp, Merkle root reference, or query origin,
  \item \( l \in L \) denotes the access level within a total access lattice \( L = \{l_1, l_2, \dots, l_n\} \) such that \( l_i \leq l_{i+1} \),
  \item \( \parallel \) denotes concatenation in the derivation domain.
\end{itemize}

The key \( K_{\mathrm{sym},i}^{l} \) is used to encrypt the knowledge fragment \( d_i \) via a symmetric cipher \( \mathcal{E} \), typically AES-256 in Galois/Counter Mode (GCM) to ensure both confidentiality and integrity:

\[
c_i = \mathcal{E}_{K_{\mathrm{sym},i}^{l}}(d_i)
\]

Each encrypted fragment is committed alongside its integrity hash and Merkle proof:

\[
f_i = \left\{ c_i, H(d_i), \text{Proof}_{M_n}(H(d_i)) \right\}
\]

The security of this schema ensures:

\begin{enumerate}
  \item Only authorised keyholders can decrypt the data.
  \item Every ciphertext is uniquely bound to its generation context.
  \item Access is privilege-tiered, preventing cross-tier key application or privilege escalation.
\end{enumerate}

Additionally, the construction admits fine-grained, fragment-level access control. That is, for every \( f_i \), there exists a unique \( K_{\mathrm{sym},i}^{l} \), disallowing any bulk key compromise from cascading across unrelated records. Moreover, no global decryption key exists—each fragment is independently sealed within its cryptographic domain.

Let \( \mathcal{D} \) be the decryption function. Then:

\[
\mathcal{D}_{K_{\mathrm{sym},i}^{l}}(c_i) = d_i \quad \text{iff} \quad K_{\mathrm{sym},i}^{l} = \text{HKDF}(K_{U,A} \parallel \text{context}_i \parallel l)
\]

The correctness condition is strictly enforced, such that decryption fails deterministically with any mismatched or insufficient privilege key.

Finally, all fragments are verified prior to decryption:

\[
\text{VerifyFragment}(H(d_i), \text{Proof}_{M_n}) = \texttt{True} \Rightarrow \text{Proceed to } \mathcal{D}_{K_{\mathrm{sym},i}^{l}}(c_i)
\]

No knowledge is accessed or reconstructed unless its historical existence and inclusion in a cryptographically committed ledger are first proven.

Hence, in this schema, symmetric encryption becomes not a layer atop memory but a structural component of it. Epistemic access is not merely authorised—it is derived. Decryption is conditional not on intent but on cryptographic merit. And so the agent does not retrieve facts. It earns them.

\section{Multi-Level Access and Ontological Privilege}

In epistemically secure systems, access is not an afterthought—it is ontologically encoded. Memory, when treated as law, must not merely resist tampering; it must actively enforce hierarchical privilege through mathematical invariants. We formalise this principle through a security lattice structure and derive access-specific decryption keys from cryptographic primitives tied to identity and context.

Let \( L = \{ l_1, l_2, \dots, l_n \} \) be a totally ordered access lattice, where each level \( l_i \) corresponds to a privilege stratum such that \( l_i \leq l_{i+1} \). Each entity \( U \) possesses a long-term elliptic curve keypair \( (sk_U, pk_U) \), and the agent \( A \) similarly holds \( (sk_A, pk_A) \).

A shared secret is established as:

\[
K_{U,A} = \text{ECDH}(sk_U, pk_A) = \text{ECDH}(sk_A, pk_U)
\]

We define a context-specific symmetric key for fragment \( i \), at privilege level \( l \), using HKDF:

\[
K_{\mathrm{sym},i}^{l} = \text{HKDF}(K_{U,A} \parallel \text{context}_i \parallel l)
\]

The derived key \( K_{\mathrm{sym},i}^{l} \) is used to decrypt only if the holder's clearance \( l_U \) satisfies:

\[
l_U \geq l
\]

In effect, key derivation becomes non-functional unless the privilege level is properly included in the derivation transcript. The privilege tag \( l \) acts as a namespace separator in the entropy space, preventing key reuse across classes.

Let the encrypted fragment be defined as:

\[
f_i = \left\{ c_i, H(d_i), \text{Proof}_{M_n}(H(d_i)) \right\}
\]

where:

\begin{itemize}
  \item \( c_i = \mathcal{E}_{K_{\mathrm{sym},i}^{l}}(d_i) \)
  \item \( H(d_i) \) is the hash of the original data payload
  \item \( \text{Proof}_{M_n} \) is a Merkle inclusion proof from memory block \( M_n \)
\end{itemize}

This construction yields three critical invariants:

\begin{enumerate}
  \item \textbf{Strict Compartmentalisation}: Each privilege level defines a disjoint keyspace. No key at level \( l_j \) can derive knowledge from \( l_k \) for \( j \neq k \).
  \item \textbf{Ontological Enforcement}: Decryption keys are not centrally stored but derived per interaction. Access is not queried—it is proven.
  \item \textbf{Forward and Backward Isolation}: Breach at one level does not propagate. An attacker with access at \( l_k \) learns nothing about \( l_{k-1} \) or \( l_{k+1} \).
\end{enumerate}

To prevent elevation by substitution, context-specific salting is mandatory. Let \( \text{context}_i \) include:

\begin{itemize}
  \item Timestamp \( t_i \)
  \item Query fingerprint \( \psi_i \)
  \item Merkle root reference \( \rho_i \)
  \item Policy contract ID \( \pi_i \)
\end{itemize}

Thus:

\[
K_{\mathrm{sym},i}^{l} = \text{HKDF}(K_{U,A} \parallel t_i \parallel \psi_i \parallel \rho_i \parallel \pi_i \parallel l)
\]

This prevents identical key reuse across contexts or data fragments. The entropy space becomes as fragmented as the data it guards.

In addition, each ontological claim is contextually bound:

\[
T_i = (s_i, p_i, o_i, H_i), \quad H_i = H(s_i \parallel p_i \parallel o_i \parallel l)
\]

where \( l \) is encoded directly into the hash. This produces privilege-scoped semantic graphs, such that ontologies themselves are non-global—an agent sees a view of reality commensurate with its clearance.

Finally, agents are embedded with privilege-level seeds \( S_A \subseteq L \). Access is computed, not assumed:

\[
\text{Access}(f_i) = 
\begin{cases}
\texttt{Permit} & \text{if } \exists l \in S_A \text{ s.t. } \mathcal{D}_{K_{\mathrm{sym},i}^{l}}(c_i) = d_i \\
\texttt{Deny} & \text{otherwise}
\end{cases}
\]

Through this construction, knowledge is stratified, not stored in the flat vector space of universal embeddings. Access is not a permission check but a cryptographic function evaluation. In this way, epistemic structures mirror the formal semantics of law: tiered, derivational, and irrevocably contextual.

\section{Immutable Learning—Preventing Catastrophic Forgetting}

Conventional machine learning paradigms are epistemologically fragile. The act of learning is tightly coupled with the act of overwriting. In such architectures, knowledge is merely the transient configuration of weights in a model—volatile, mutable, and pathologically prone to catastrophic forgetting. This phenomenon, in which previously acquired capabilities are lost during the integration of new knowledge, arises from the absence of structural memory guarantees. Our alternative is an immutable learning paradigm grounded in append-only, cryptographically verifiable graph structures.

We define a directed acyclic graph \( G = (V, E) \), where:

\[
V = \{ K_1, K_2, \ldots, K_n \}, \quad E = \{ (K_i, K_j) \mid K_j \text{ refines or extends } K_i \}
\]

Each node \( K_i \in V \) represents a committed fragment of knowledge, structured as:

\[
K_i = \left\{ d_i, p_i, H(d_i \parallel p_i), \mathcal{E}_{k_i}(d_i), \text{Proof}_{M_i}(H(d_i)) \right\}
\]

Here:
\begin{itemize}
  \item \( d_i \) is the data content,
  \item \( p_i \) denotes provenance metadata (e.g., timestamp, author, context),
  \item \( H \) is a collision-resistant hash function (e.g., SHA-256),
  \item \( \mathcal{E}_{k_i}(d_i) \) denotes encryption under a symmetric key derived from an ECDH-HKDF pipeline,
  \item \( \text{Proof}_{M_i} \) is a Merkle inclusion proof for auditability.
\end{itemize}

Knowledge fragments are never deleted. Updates manifest as new nodes \( K_j \) with explicit refinement edges \( (K_i, K_j) \). The refinement relation induces a topological constraint that enforces semantic continuity: no knowledge claim stands unreferenced or contextless.

\subsection*{Delta Encoding of Knowledge Transitions}

Updates are encoded as deltas. Let \( \Delta(K_i, K_j) \) denote the semantic difference between two fragments. The system records only \( \Delta \), and \( K_j \) inherits \( K_i \)'s reference path. This compression ensures efficient storage while maintaining traceability. Formally:

\[
K_j = \text{Apply}(K_i, \Delta(K_i, K_j))
\]

\[
\Delta(K_i, K_j) = \text{Diff}(K_i, K_j)
\]

Since \( \text{Apply} \) and \( \text{Diff} \) are invertible and deterministic, the full knowledge lineage is recoverable.

\subsection*{Topological Constraints and Acyclicity}

The graph \( G \) must remain acyclic. Formally:

\[
\forall K_i, K_j \in V, \quad (K_j, K_i) \notin E \quad \text{if } (K_i, K_j) \in E
\]

Cycles are epistemologically invalid—no fragment may depend on its refinement. This restriction preserves historical integrity and guards against retroactive alteration.

\subsection*{Knowledge Lineage Proofs}

Given the root node \( K_0 \), the full ancestry of any node \( K_n \) is verifiable via a path:

\[
P = \{ K_0, K_1, \ldots, K_n \} \quad \text{such that } (K_{i-1}, K_i) \in E
\]

Let \( \rho = \text{MerkleRoot}(\{ H(K_i) \}) \). The lineage is verifiable using the inclusion proof from \( \rho \), independent of trusting any particular agent. Thus:

\[
\text{Verify}(K_n, \rho, \text{Proof}_P) = \texttt{True}
\]

Only fragments with complete and valid chains are admissible in inference.

\subsection*{Training as Graph Extension}

Training does not mutate parameters; it produces graph extensions. A knowledge acquisition step maps:

\[
\text{Learn} : K_i \mapsto K_j \quad \text{with } (K_i, K_j) \in E
\]

The model is constrained to reason over the closure \( \text{cl}(K_j) \), defined as the set of all fragments reachable from \( K_j \) via backward traversal. This closure forms the model’s epistemic context. Thus, epistemic drift is structurally impossible.

\subsection*{Immutable Evaluation Contexts}

Evaluation is scoped to explicit subgraphs. Let \( \Sigma \subseteq V \) be a subset of knowledge nodes. Inference proceeds over \( \Sigma \) only if:

\[
\forall K_i \in \Sigma, \quad \exists \text{Proof}_{M_i}(H(K_i))
\]

This cryptographic gate ensures that only valid, committed fragments participate in reasoning. Spurious or ephemeral data is categorically excluded.

\subsection*{Agent Behaviour Constrained by Provenance}

The system enforces:

\[
\text{Output}_A(t) = f \Rightarrow \exists K_i \in V \text{ such that } f \in \text{Closure}(K_i)
\]

In other words, an agent may only emit outputs that are provably grounded in the knowledge graph. This eliminates hallucination by fiat.

\subsection*{Implications for Model Design}

This framework obliterates the notion of model versioning as a sequence of overwritten states. Instead, each model state is a snapshot of a verifiable subgraph. Model evaluation becomes a function:

\[
\mathcal{M} : \text{Input} \times \text{GraphSnapshot} \rightarrow \text{Output}
\]

Consequently, the same model binary may produce different outputs when paired with different epistemic contexts—without violating determinism or auditability.

\subsection*{Summary}

Catastrophic forgetting is a symptom of mutable epistemologies. The resolution is not continual fine-tuning but architectural reformation. Immutable learning reifies knowledge as committed, referentially complete, and historically situated. By encoding knowledge transitions as acyclic, cryptographically linked deltas, the system elevates inference from an exercise in approximation to a constrained traversal of verifiable history. In this, the machine ceases to learn as we do—not by amnesia—but by memory that cannot forget.

\section{Memory as "Law"—Cryptographic Truth and Legislative Ontologies}

Artificial agents built on mutable memory risk devolving into instruments of persuasion rather than structures of truth. In such agents, knowledge becomes fungible; assertions are updated without lineage, and justification is reduced to statistical correlation. This is not epistemology—it is rhetorical modelling. We reject this paradigm in favour of a formal system where knowledge is stored and reasoned over as law: immutable, referential, and structured by legislative logic.

To formalise this, memory must obey three constraints:
\begin{enumerate}
    \item Provenance: every fact must be sourced, timestamped, and anchored.
    \item Immutability: past facts cannot be erased or rewritten.
    \item Lineage: every inference must trace to axiomatic foundations.
\end{enumerate}

We define the fragment structure \( T_i \) as:

\[
T_i = (s_i, p_i, o_i, H_i), \quad \text{where } H_i = H(s_i \parallel p_i \parallel o_i)
\]

Each triple \((s_i, p_i, o_i)\) is a Resource Description Framework (RDF) statement representing subject, predicate, and object. The cryptographic hash \( H_i \) anchors the triple into a Merkle tree \( M_n \), which is itself committed to a blockchain ledger via:

\[
\text{Root}(M_n) \hookrightarrow \text{Tx}(B_n)
\]

Thus, the ontology becomes historically sealed. No claim can be made unless its triple exists and is verifiably on-chain.

\subsection*{Legislative Semantics via Description Logics}

We model ontologies using DL-Lite and OWL 2, which support tractable reasoning while maintaining logical expressivity. Each class assertion, property assertion, and constraint is recorded as an anchored fragment. For example:

\[
\text{ClassAssertion}(Professor, \text{CraigWright}) \mapsto T_i
\]

\[
\text{SubClassOf}(Cryptographer, Scientist) \mapsto T_j
\]

Every axiom is encoded as an immutable triple and anchored. Logical inference is now constrained by on-chain ontological entailments. That is, the AI may only conclude \( C \sqsubseteq D \) if a provable chain of subclass relations exists, each committed on-chain.

\subsection*{Epistemic Reasoning as Jurisprudential Traversal}

Let the knowledge base be \( \mathcal{K} = \{ T_1, \ldots, T_n \} \). We define a reasoning query \( q \) as admissible iff:

\[
\exists \pi = \{ T_{i_1}, T_{i_2}, \ldots, T_{i_k} \} \subseteq \mathcal{K} \quad \text{such that } \pi \models q
\]

Where \( \pi \models q \) is provable under a description logic reasoner that only operates on fragments with valid inclusion proofs. If \( \text{VerifyProof}_{M_j}(H(T_{i_l})) = \texttt{True} \) for all \( l \in [1,k] \), the result is sanctioned. Otherwise, it is void ab initio.

\subsection*{Regulatory Codes as Root Ontologies}

We further classify certain ontologies as regulatory codes. These include ethics guidelines, contractual obligations, and legal statutes. Each code is stored as a named graph \( \mathcal{G}_r \subset \mathcal{K} \) with:

\[
\mathcal{G}_r = \{ T_i \mid \text{jurisdiction}(T_i) = r \}
\]

The agent's admissible actions are now subject to:

\[
\forall a \in \text{Actions}, \quad \text{Permissible}(a) \iff \mathcal{G}_r \models \text{Permitted}(a)
\]

This enforces behavioural constraints as legal entailments. AI actions are no longer probabilistic outcomes but legislative consequences.

\subsection*{Immutable Legal Precedent}

We define legal precedent not as analogical similarity but as committed entailment chains. Let \( P = \{ T_a, T_b, T_c \} \) be a precedent sequence. Then:

\[
\text{NewAction}(A) \text{ is valid} \iff P \models \text{Permitted}(A)
\]

Agents do not speculate—they appeal to immutable fragments with verifiable provenance. This transforms the act of inference into jurisprudential citation.

\subsection*{Final Implications}

In this architecture, the knowledge base of an agent functions not as a fluid vector space but as a legislative corpus. Ontological terms are interpreted within anchored vocabularies; reasoning is constrained by logic; and actions are judged not by consequence, but by legality.

Truth, in this regime, is not a belief state.

It is a provable, cryptographically grounded claim.

\section{Epilogue—To Build Memory Without Mind}

The prevailing trajectory of artificial intelligence research has been largely defined by attempts to emulate human cognition. Whether through neural approximations or probabilistic generative models, the ambition to recreate a facsimile of consciousness remains dominant. However, such efforts conflate epistemic reliability with cognitive simulation. The human mind—rife with bias, error, and inconsistent recall—is not a suitable archetype for systems intended to serve as objective computational agents. Emulating the mind is not only unnecessary, but epistemologically unsound when the goal is verifiable, immutable knowledge representation.

This work proposes an alternative paradigm: the construction of an epistemic substrate in which memory is not merely functional but jurisprudential. Here, memory is conceived not as an act of retrieval, but as an artefact of commitment—an immutable, cryptographically sealed ledger of transitions, assertions, and derivations. Each knowledge fragment is not merely a parameter or vector, but a historically anchored datum whose origin, integrity, and access are mathematically constrained. The system does not rely on recall from learned weights but performs traversal over a Merkle-anchored state space.

In this construction, cognition is operationalised as execution through a verified state transition system, constrained by formal ontologies and subject to access policies defined cryptographically. Each assertion is validated through Merkle inclusion proofs, each inference formed through DAG-consistent lineage, and each behavioural output regulated through precommitted epistemic contracts. In doing so, the architecture resists the principal failure mode of current generative systems: hallucination.

Furthermore, the system decouples belief from truth. It is agnostic to perception and affect; it neither speculates nor imagines. Rather, it operates within a formal system wherein each output is logically and cryptographically deducible from prior commitments. Provenance is enforced through digital signatures and block-level timestamping; revisionism is rendered computationally infeasible. Knowledge cannot be altered ex post without cryptographic contradiction.

This framework advances the thesis that epistemic reliability in artificial agents is achievable not through greater intelligence but through more rigorous constraint. The value of such a system lies not in its capacity to mimic human faculties but in its incapacity to violate its own formal foundations. Trust emerges not from persuasion or affective resonance but from structural guarantees, cryptographic enforcement, and logical closure.

In conclusion, the future of artificial agency does not require consciousness, affect, or the illusions of sentience. It requires systems that can commit, systems that cannot forget, and systems that cannot fabricate. The architecture described herein establishes the foundation for such agents—synthetic witnesses whose utility derives from their fidelity to constraint, not their resemblance to biological minds.

Thus, we argue that artificial memory should be constructed not as a metaphorical mirror of human thought, but as a formal, verifiable institution of computational truth. In rejecting the psychological model of cognition, we embrace a more principled framework—one in which epistemology becomes executable and truth becomes a function of structure, not belief.

\section{Additional Structures: Append-Only Reasoning Graphs}

While the preceding sections outline a foundation for immutable memory through cryptographically committed automata and ontological constraint, the expansion of agentic reasoning requires a structural formalism that permits evolution without revision. To that end, we define the Append-Only Reasoning Graph (AORG), a directed acyclic graph (DAG) of logical inference nodes where each node corresponds to a committed knowledge claim, derivation, or computation, and each edge represents a cryptographically verifiable dependency.

Let \( G = (V, E) \) denote the AORG, where \( V = \{ v_1, v_2, \ldots, v_n \} \) are vertices representing discrete reasoning events, and \( E \subseteq V \times V \) are directed edges such that if \( (v_i, v_j) \in E \), then \( v_j \) is a logical or inferential refinement of \( v_i \). Each vertex is formally defined as:

\[
v_i = \left( \phi_i, \sigma_i, h_i, t_i \right)
\]

where:
\begin{itemize}
  \item \( \phi_i \) is the formal expression or proposition at node \( i \),
  \item \( \sigma_i \) is the signature (digital, multi-party, or threshold) verifying the author or authority of the inference,
  \item \( h_i = H(\phi_i \parallel \sigma_i \parallel t_i) \) is the cryptographic hash of the node content, and
  \item \( t_i \) is the monotonic timestamp committed at a block height or ledger position.
\end{itemize}

Edges \( (v_i, v_j) \) in this structure must be justified by formal inference rules, such that the proposition \( \phi_j \) is either a direct logical consequence of \( \phi_i \) under a known system (e.g., natural deduction, sequent calculus), or a refinement under ontological constraint. No edge may be cyclically introduced; that is, \( G \) must preserve acyclicity to maintain directional epistemic growth.

Each new addition to the graph is subjected to a set of verifiability constraints:
\begin{enumerate}
  \item \textbf{Consistency}: \( \phi_j \) must not contradict any ancestor node \( \phi_k \) in its transitive closure.
  \item \textbf{Justifiability}: Each edge must correspond to an application of a rule in the formal reasoning system \( R \).
  \item \textbf{Finality}: Once committed and anchored, no vertex may be deleted or edited; knowledge growth is strictly append-only.
  \item \textbf{Traceability}: Any \( \phi_j \) must be reducible via its path to initial axiomatic or empirical root nodes.
\end{enumerate}

The AORG constitutes a memory and inference substrate simultaneously. Unlike static knowledge graphs, it includes not only declarative facts but inferential steps, epistemic obligations, and revision-resistant justifications. This makes it possible to construct verifiable proofs-of-thought, where the derivation of an output or belief is not reconstructed heuristically, but followed as a certified computational trace.

Critically, the append-only nature of the structure prevents retrospective manipulation or deletion of epistemic commitments. This eliminates re-training biases, temporal inconsistencies, and the "forgetting" of past states. Each claim once added to the graph is a matter of permanent record and any contradiction must be issued as a refinement node, forming a fork with higher evidential or inferential authority.

Finally, reasoning graphs can be versioned and scoped. Let \( G_i^k \) denote a subgraph rooted in \( v_i \) with a derivational depth of \( k \). This allows for modular provenance, memory partitioning, and domain-specific inferential agents operating over bounded reasoning windows.

In sum, the AORG furnishes the agent with a memory model that is also a formal logic engine, binding cognitive growth to mathematical and cryptographic invariants. Such agents do not merely store facts—they prove them, trace them, and are bound by their own reasoning history. The agent becomes not a stochastic regurgitator of patterns but a witness whose statements are theorems in a committed, verifiable epistemic calculus.

\section{Clock Anchoring and Consensus-Time Inference}

In decentralised epistemic architectures, the passage of time must not rely on local system clocks susceptible to drift, rollback, or manipulation. Instead, we define consensus-time as an externally verifiable, cryptographically enforced ordering mechanism anchored in blockchain data structures. The objective is to enable synthetic agents to reason about time with the same formality and auditability applied to logic and memory.

Let \( T = \{ t_1, t_2, \ldots, t_n \} \) denote a discrete set of temporal anchors, where each \( t_i \) corresponds to a blockchain block header timestamp \( \tau_i \), committed in the Merkle root of block \( B_i \). While \( \tau_i \) is not itself trustworthy, its position in the chain enforces partial ordering: \( B_i < B_{i+1} \Rightarrow \tau_i \leq \tau_{i+1} \). We define consensus time \( \mathbb{T}_C \) as the monotonic function:

\[
\mathbb{T}_C : \mathbb{N} \rightarrow \mathbb{R}, \quad \mathbb{T}_C(i) = \tau_i
\]

Where the index \( i \) corresponds to the blockchain height and \( \tau_i \) is the timestamp associated with block \( i \), adjusted by protocol rules to constrain forward skew and manipulation (e.g., median-of-past-N rules in Nakamoto consensus systems).

Each memory fragment or reasoning node \( v_k \) in the agent's append-only reasoning graph (AORG) is annotated with a consensus timestamp:

\[
v_k = \left( \phi_k, \sigma_k, h_k, \mathbb{T}_C(i_k) \right)
\]

This ensures that epistemic commitments are ordered not by system time but by canonical block height and timestamp pairs. Let \( \Delta(v_j, v_k) = \mathbb{T}_C(i_j) - \mathbb{T}_C(i_k) \). This difference is verifiable and non-subjective, anchoring relative inference intervals in ledger-based truth rather than mutable internal clocks.

In cases where multiple blockchains are consulted (e.g., inter-chain anchoring or cross-ledger commitment), the agent employs a consensus-time aggregation function:

\[
\mathbb{T}_C^{*}(i) = f\left( \mathbb{T}_{C_1}(i_1), \ldots, \mathbb{T}_{C_m}(i_m) \right)
\]

Where \( f \) is an aggregation operator such as weighted median or cryptographic quorum timestamping across chains \( C_1 \ldots C_m \). This permits agents to resolve disputed clocks through externalised consensus rather than internal arbitration.

To enable temporal logic, the reasoning system is extended with operators over consensus-time:

\begin{itemize}
  \item \( \Box_t \phi \): it is always the case that \( \phi \) holds at or before consensus-time \( t \),
  \item \( \Diamond_{(t_1, t_2)} \phi \): \( \phi \) holds at some time between \( t_1 \) and \( t_2 \),
  \item \( \phi \rightarrow_t \psi \): if \( \phi \) holds at \( t \), then \( \psi \) must follow within \( \Delta t \).
\end{itemize}

These temporal modalities enable the construction of time-aware proofs, schedules, SLAs, and legal obligations, transforming the agent into a chronologically accountable witness.

Furthermore, temporal forks or contradictory sequences are resolved by dominance in the longest chain or pre-specified trust anchors. No reasoning path may cite events after their committed consensus-time, enforcing causal closure.

In conclusion, clock anchoring provides more than time ordering. It yields a foundation for accountability, causality, and audit across all inferential acts. Through cryptographic enforcement of monotonic temporal evolution, agents no longer rely on fragile clock signals or system heuristics. They reason in time the way they reason in logic: by proof, by constraint, and by unambiguous, externalised order.

\section{Zero-Knowledge Inclusion Proofs for Memory Access}

In epistemically constrained systems, access to memory must be provable without disclosure. This creates a fundamental tension: the agent must reveal neither its internal state nor the data requested, while simultaneously proving possession and proper derivation. We resolve this via the integration of zero-knowledge proofs (ZKPs), specifically structured for Merkle-inclusion verification.

Let the agent’s memory be committed in a Merkle tree \( M_n \), rooted in \( \mathcal{R}_n \), with leaves \( L = \{ h_1, h_2, \ldots, h_k \} \), where each \( h_i = H(d_i) \) is the cryptographic hash of memory fragment \( d_i \). Suppose an external verifier \( V \) challenges the agent \( A \) to prove access to \( d_j \) without revealing \( d_j \) itself.

Define:

\[
\pi_j = \text{ZKMerkleProve}(h_j, \mathcal{R}_n, \mathcal{P}_j)
\]

Where \( \mathcal{P}_j \) is the authentication path from \( h_j \) to \( \mathcal{R}_n \), and \( \pi_j \) is a zero-knowledge proof of inclusion. The protocol ensures that:

\begin{enumerate}
    \item \( \pi_j \) is succinct and non-interactive (via Fiat–Shamir heuristic),
    \item \( V \) can verify \( h_j \in M_n \) without learning \( d_j \),
    \item \( A \) cannot fabricate membership of a non-existent \( h_j \) due to collision-resistance of \( H \).
\end{enumerate}

For secure multi-party access, the proof is extended with access control tags. Let each fragment \( d_j \) be encrypted as \( f_j = \{ \text{Enc}_{K_{\text{sym},j}}(d_j), H(d_j) \} \), with the derived key \( K_{\text{sym},j} \) tied to a privilege level \( \ell_j \). The agent provides:

\[
\Pi_j = \left( \pi_j, \text{ZKAccess}(\ell_q \geq \ell_j) \right)
\]

Where \( \text{ZKAccess} \) proves that the querying entity’s access level \( \ell_q \) satisfies \( \ell_q \geq \ell_j \) without revealing either party’s full credentials. This creates a dual commitment: one to the existence of the data and one to the legitimacy of the access request.

In dynamic environments, where memory is updated incrementally, a forward-secure proof chain is established. Every new Merkle root \( \mathcal{R}_{n+1} \) commits to the prior root:

\[
\mathcal{R}_{n+1} = H(\mathcal{R}_n \parallel \Delta M_{n+1})
\]

Thus, ZK-inclusion proofs for any fragment \( d_j \in M_t \) can be extended to \( M_{n} \), \( n \geq t \), by recursive anchoring. Proofs remain valid across agent evolution, enabling immutable and portable attestations.

For broader interoperability, we define a formal language \( \mathcal{L}_{ZKMem} \) in which inclusion proofs are encoded as elements of a decidable grammar:

\[
\mathcal{L}_{ZKMem} = \{ \Pi_j \mid \Pi_j \vdash (h_j \in M_n) \land (\ell_q \geq \ell_j) \}
\]

This allows agents to exchange memory access proofs across domains and validate access histories using standard ZKP verifiers. No central authority is required—only a consensus ledger and public parameters for \( H \), \( \text{ZKProve} \), and \( \text{ZKAccess} \).

In conclusion, zero-knowledge inclusion proofs enable epistemic integrity without informational leakage. Memory becomes a sealed cryptographic artefact—provable, referential, and privately accessible. The agent becomes not a vault of secrets but a structure of proofs: nothing is seen, but everything is verified.

\section{Policy Derivation as Deductive Graph Traversal}

In computational systems tasked with regulatory compliance or legal reasoning, policy adherence must be verifiable, auditable, and non-heuristic. Traditional approaches embed policy within code logic—opaque, brittle, and semantically shallow. We assert instead that policies must be encoded as formal deductive structures: graphs composed of verifiable premises and consequence rules, embedded within an immutable substrate. This reifies policy not as a document but as a directed acyclic graph (DAG) of derivable constraints.

Let the policy ontology be defined as a graph \( G = (V, E) \), where each node \( v_i \in V \) is a formally grounded axiom, rule, or decision clause, and each directed edge \( (v_i, v_j) \in E \) represents a deductive dependency: that is, \( v_i \rightarrow v_j \) if \( v_j \) is logically derivable from \( v_i \). Let \( \mathcal{L}_{\text{pol}} \) be the formal language governing the syntax of permissible statements. All vertices must be well-formed formulae (WFFs) in \( \mathcal{L}_{\text{pol}} \), and edges must conform to a verified rule schema, denoted \( \vdash_{\mathcal{R}} \), where:

\[
(v_i, v_j) \in E \iff v_i \vdash_{\mathcal{R}} v_j
\]

To anchor such graphs, each node \( v_i \) is cryptographically committed via:

\[
h_i = H(\text{serialise}(v_i))
\]

and each edge is committed as a tuple:

\[
e_{ij} = (h_i, h_j, \tau_{ij}, H(h_i \parallel h_j \parallel \tau_{ij}))
\]

where \( \tau_{ij} \) encodes the specific rule applied in the deduction (e.g., Modus Ponens, Universal Instantiation). These are included in a Merkle-anchored DAG structure, such that the policy graph root commits to the entire reasoning structure.

Policy queries are resolved by path discovery: given a terminal conclusion \( v_k \), the agent must construct a valid derivation path \( \pi_k = \{v_0, v_1, \dots, v_k\} \) such that each transition satisfies \( \vdash_{\mathcal{R}} \), and the entire path is anchored and cryptographically verifiable. Inclusion proofs are provided using Merkle paths over the DAG, along with zero-knowledge derivation attestations where required.

In systems with stratified policy domains (e.g., finance, data protection, civil rights), we define a layered structure of graphs \( \{G_1, G_2, \ldots, G_n\} \), with formal mappings:

\[
\Phi_{i \rightarrow j}: V_i \rightarrow V_j
\]

to support inheritance and contextual overrides. For example, national data privacy policies (e.g., GDPR) may instantiate core principles from supranational human rights frameworks via such morphisms.

This layered DAG structure allows for efficient contradiction detection. Suppose \( v_i \in G_x \) and \( v_j \in G_y \) with \( v_i \rightarrow \neg v_j \), and a path \( \pi \) exists such that both \( v_i \) and \( v_j \) are derivable. Then:

\[
\text{Conflict}(\pi) = \exists (v_i, v_j) \text{ such that } \pi \vdash v_i \land \pi \vdash v_j \land v_i \rightarrow \neg v_j
\]

This enables agents to not only reason within a policy system but to identify and localise violations, inconsistencies, or overrides.

Finally, to ensure updatability without erasure, policy graphs evolve via append-only revisions. Each new graph version \( G^{(t+1)} \) includes cryptographic linkage to \( G^{(t)} \), and any overridden nodes carry a reference to their superseded version. This temporal graph chain enforces historical integrity while allowing legal evolution:

\[
G^{(t+1)} = G^{(t)} \cup \Delta G^{(t+1)}, \quad \text{with } \forall v' \in \Delta G^{(t+1)}, \exists v \in G^{(t)} : v' \succ v
\]

Where \( \succ \) denotes policy succession, provable through formal diff and proof annotation. The result is a system in which policy is neither a document nor code, but an epistemic graph: self-referential, rule-governed, and cryptographically immutable.

In such a framework, synthetic agents do not apply policy heuristically—they deduce permitted actions via traversal. They are not interpreters of law; they are traversers of law’s proof graph. In every decision, the trace of logic is preserved, sealed, and referentially intact.

\section{Failure Modes: Inconsistency, Contradiction, and Revision}

Even within cryptographically grounded epistemic structures, failure remains possible—and in a system of immutable commitments, failure bears permanence. This section characterises epistemological failure modes within append-only knowledge architectures, each tied not to stochastic behaviour, but to logical defect and procedural omission.

Let the reasoning substrate be formalised as a directed acyclic graph \( G = (V, E) \), where each vertex \( \phi_i \in V \) is a committed knowledge fragment (either axiomatic or derived), and each edge \( e_{i,j} \in E \) represents a deductive inference from \( \phi_i \) to \( \phi_j \). The integrity of the structure depends upon both local and global coherence.

\subsection*{Types of Failure}

\begin{enumerate}
  \item \textbf{Inconsistency}: A fragment \( \phi_j \) is inconsistent if there exists a pair \( (\phi_k, \phi_m) \in \text{anc}(\phi_j) \) such that \( \phi_k \vdash \psi \) and \( \phi_m \vdash \neg \psi \) for some proposition \( \psi \). Even if each inference step is locally valid, contradiction at the transitive level poisons epistemic certainty.

  \item \textbf{Contradiction}: When \( \exists \phi_i, \phi_j \in V \) such that \( \phi_i = \neg \phi_j \), and both are reachable from a shared root \( \phi_0 \), the DAG commits a contradiction. This cannot be resolved by erasure. As the structure is append-only, contradiction implies the epistemic root is corrupted or the inference path is ill-formed.

  \item \textbf{Invalid Justification}: An edge \( e_{i,j} \) is illegitimate if the transition from \( \phi_i \) to \( \phi_j \) cannot be reconstructed under the formal system \( R \). That is, \( \nexists r \in R \) such that \( \phi_i \xRightarrow{r} \phi_j \). These correspond to hallucinations—not in the statistical sense, but in violation of deductive closure.

  \item \textbf{Revision via Accretion}: An agent may attempt to overwrite \( \phi_j \) by adding a successor \( \phi_{j'} \) with a contradictory statement, simulating revision. However, because all paths remain committed, this introduces bifurcation rather than correction. A consistent DAG cannot accommodate both \( \phi_j \) and \( \phi_{j'} = \neg \phi_j \) unless explicit modality is used.

  \item \textbf{Contextual Drift}: Fragments whose interpretation depends on temporal, legal, or ontological context may be misaligned as upstream referents change. Let \( \phi_j = \text{“Act A is legal under Reg X”} \). If Reg X is revised, the truth-value of \( \phi_j \) is historically correct but presently misleading. Without contextual anchoring (e.g., timestamps, source hashes), reasoning inherits silent anachronism.
\end{enumerate}

\subsection*{Mitigations}

To defend against these failure modes, the following structural safeguards are imposed:

\begin{enumerate}
  \item \textbf{Consistency Constraints}: DAG construction enforces that any new \( \phi_j \) must not contradict any ancestor node \( \phi_k \) in its transitive closure. That is, before committing \( \phi_j \), verify that \( \forall \phi_k \in \text{anc}(\phi_j), \neg (\phi_k \vdash \neg \phi_j) \).

  \item \textbf{Justification Verification}: Each edge must correspond to a rule in \( R \), and must include a cryptographically committed proof sketch \( \pi_{i,j} \) of the deduction \( \phi_i \vdash_R \phi_j \). This enables on-chain audit of logical provenance.

  \item \textbf{Finality Guarantees}: Once committed and anchored, no vertex may be deleted or edited; knowledge growth is strictly append-only. Corrections must be issued as new nodes with explicit references and contextual disclaimers.

  \item \textbf{Traceability Enforcement}: Any \( \phi_j \) must be reducible via its path to initial axiomatic or empirical root nodes. If not, the node is rejected as unverifiable.

  \item \textbf{Modality Encoding}: To handle contradictions arising from legal or temporal change, fragments include modal qualifiers—deontic, epistemic, or temporal—that are themselves ontologically committed. E.g., “as of Block 105432, Reg X includes Clause Y.”
\end{enumerate}

Failure in such a system is not erased—it is merely contextualised. Truth is not updated; it is extended. That which was wrong remains visible, and that which is corrected is anchored as a response—not a replacement. In this sense, epistemic failure becomes part of the archive, not the erasure of it.

\section{Audit Trails and Agent Liability in Causal Chains}

In systems where artificial agents act as autonomous decision-makers, accountability hinges not on subjective intent but on reconstructible causal chains. This section defines the formal structure of audit trails in cryptographically committed automata and establishes liability through deterministic traceability.

Let an agent \( A \) be modelled as a state-transition system with ledger-bound memory, executing a sequence of decisions \( D = \{ d_1, d_2, \dots, d_n \} \), where each decision \( d_i \) results from a computation over inputs \( \mathcal{I}_i \), internal state \( s_i \), and a committed knowledge graph \( G_i \). Formally:

\[
d_i = \mathcal{F}(s_i, \mathcal{I}_i, G_i), \quad \text{with } \mathcal{F} \text{ deterministic and reproducible}
\]

Each decision is committed as a tuple:

\[
T_i = \left( d_i, H(s_i), H(\mathcal{I}_i), H(G_i), \tau_i \right)
\]

where \( H(\cdot) \) denotes a collision-resistant hash function, and \( \tau_i \) is the timestamp of the decision’s commitment. These tuples are sequentially anchored into a Merkle chain, such that:

\[
\text{Root}_n = \text{Merkle}(T_1, T_2, \dots, T_n)
\]

Liability is thus defined not through legal personhood, but through reconstructible inference. If harm \( \mathcal{H} \) arises from \( d_k \), one queries the chain of \( T_i \) up to \( T_k \), verifying that:

\begin{enumerate}
  \item The state \( s_k \) is reachable via the transition function from \( s_0 \) and all preceding \( d_i \), i.e., \( s_k = \delta(s_0, d_1, \dots, d_{k-1}) \)
  \item The knowledge graph \( G_k \) reflects only fragments to which the agent had verifiable access, i.e., \( G_k \subseteq \{ \phi_i \mid \exists \pi_i : \phi_i \in \text{Proof}_{M_n}(H(\phi_i)) \} \)
  \item The decision function \( \mathcal{F} \) conforms to its formal specification and is provably deterministic
\end{enumerate}

If any of these conditions fail, agent behaviour becomes non-compliant, and liability transfers to the operating entity. Importantly, when the agent is cryptographically sealed, audit trails become immune to post hoc manipulation. Thus, we redefine culpability as a function of verifiable procedure, not interpretive motive.

We now formalise the causality trail \( \mathcal{C}(d_k) \) as:

\[
\mathcal{C}(d_k) = \{ (T_j, r_j) \mid j \leq k, \; r_j \text{ encodes the inferential or causal role of } T_j \text{ in producing } d_k \}
\]

Each \( r_j \) is a logical or functional descriptor (e.g., “premise,” “inference step,” “conditional branch,” etc.). Together, \( \mathcal{C}(d_k) \) provides a minimal reconstructive certificate of the decision’s epistemic provenance.

Finally, agent liability must account for delegation. Let \( A \) receive sub-decisions from \( A' \). Then \( T_i \) includes a delegation map:

\[
\Delta_i = \{ (A', d'_j, \sigma_j) \}
\]

where \( \sigma_j \) is a signature from \( A' \) asserting authorship over \( d'_j \). This ensures that multi-agent systems retain explicit, signed provenance for every causal link, enabling attribution across organisational boundaries.

In such architectures, audit becomes forensic mathematics. Liability becomes a deductive function. Accountability no longer depends on human memory or testimony, but on cryptographic evidence chains embedded into the system’s temporal substrate.

\section{Modular DAG Instancing for Partial Context Awareness}

In complex artificial reasoning systems, full context loading is computationally prohibitive and epistemically unnecessary. Instead, agents may instantiate partial subgraphs of a committed global DAG (Directed Acyclic Graph) representing all known knowledge. This section defines the principles and formal conditions under which such partial instances maintain consistency, determinism, and bounded rationality.

Let the global knowledge structure be a DAG \( G = (V, E) \), where each node \( v_i \in V \) corresponds to a knowledge fragment \( K_i \) and each directed edge \( e_{ij} = (v_i, v_j) \in E \) denotes that \( K_j \) is derivable from \( K_i \). The entire structure is cryptographically anchored such that:

\[
\text{Root}_G = \text{Merkle}(\{ H(K_i) \mid v_i \in V \})
\]

A reasoning agent does not operate on \( G \) in full. Rather, it instantiates a modular subgraph \( G' = (V', E') \subset G \) such that \( V' \subseteq V \), \( E' \subseteq E \cap (V' \times V') \), and \( G' \) satisfies closure under derivability:

\[
\forall v_j \in V', \; \text{if } (v_i, v_j) \in E, \text{ then } v_i \in V'
\]

This guarantees that no inferred statement appears without its prerequisite axioms or propositions. The subgraph is thus deductively coherent even in isolation.

Let \( \Gamma \subseteq V' \) be the active working set for a query \( q \). The reasoning engine defines a partial entailment structure:

\[
\Gamma \vdash_q \phi \quad \text{iff } \phi \in \text{Closure}(\Gamma, R_q)
\]

where \( R_q \) is the rule subset relevant to the scope of \( q \). This restricts inferential breadth to only those rules compatible with the instantiated fragment.

To prevent contextual contradictions, subgraphs are required to satisfy Merkle consistency proofs. That is, for any \( v_i \in V' \), the agent provides \( \text{Proof}_{M_G}(H(K_i)) \), a Merkle path demonstrating that the fragment is indeed part of the canonical DAG. This removes the risk of adversarial insertion or hallucination of knowledge.

Furthermore, let \( C_i = \{ c_1, c_2, \dots \} \) be a context vector attached to each node \( v_i \). These context vectors are used for query-relevant instancing:

\[
G' = \text{Instance}(G, \Psi) = \left( V', E' \right) \quad \text{where } V' = \{ v_i \in V \mid \Psi \cap C_i \neq \emptyset \}
\]

Here, \( \Psi \) is the predicate context—e.g., legal domain, temporal constraints, or geographical applicability. This ensures that even partial context loads remain structurally and semantically scoped.

In multi-agent settings, modular DAGs enable scoped delegation. Let \( A \) be the querying agent and \( A' \) the delegating entity. Then, a delegated graph instance \( G'_{A'} \) comes with signature \( \sigma_{A'} \) attesting to the validity of the instantiation function and content set:

\[
\sigma_{A'} = \text{Sign}_{sk_{A'}}(H(G'_{A'} \| \Psi \| \tau))
\]

where \( \tau \) is a timestamp sealing the instance. Verification includes both content and scope, binding responsibility for inference to the delegator.

Thus, modular DAG instancing allows scalable, efficient, and legally bounded reasoning. It introduces a principled mechanism to traverse only what is required—no more, no less—while guaranteeing the integrity and legitimacy of the agent’s contextual awareness.

\section{Formal Verification of DAG Constraints in Agent Outputs}

In architectures anchored to immutable knowledge graphs, formal verification of agent outputs becomes a foundational necessity. Given a committed DAG \( G = (V, E) \), where nodes represent propositional fragments and edges encode derivational relationships, the outputs \( \Phi = \{\phi_1, \phi_2, \dots, \phi_n\} \) of any reasoning agent must satisfy structural, logical, and cryptographic constraints derivable from \( G \). This section outlines a formal model to verify such constraints.

Let each output \( \phi_i \in \Phi \) be associated with a derivation trace \( T_i = (v_{i_1}, v_{i_2}, \dots, v_{i_k}) \), where \( v_{i_k} \in V \) and \( v_{i_j} \rightarrow v_{i_{j+1}} \in E \) for all \( j < k \), culminating in \( \phi_i \). Then, the following properties must hold:

\begin{enumerate}
  \item \textbf{Closure Under Inference:} For each \( T_i \), the inference \( \phi_i \) must be valid under the deductive system \( R \), i.e.,
  \[
  \phi_i \in \text{Closure}(\{ K(v_{i_1}), \dots, K(v_{i_{k-1}}) \}, R)
  \]
  where \( K(v_j) \) denotes the knowledge fragment committed at node \( v_j \).

  \item \textbf{Lineage Integrity:} For each \( v_j \in T_i \), a Merkle inclusion proof \( \text{Proof}_{M_G}(H(K(v_j))) \) must exist, ensuring its anchoring in the canonical DAG. Formally:
  \[
  \forall v_j \in T_i, \; \exists \; \pi_j : \text{VerifyMerkle}(\pi_j, H(K(v_j)), \text{Root}_G) = \text{true}
  \]

  \item \textbf{Cycle Prohibition:} Since \( G \) is a DAG, \( T_i \) must not revisit any \( v_j \). The verifier must check:
  \[
  \forall v_j, v_k \in T_i, \; j \neq k \Rightarrow v_j \neq v_k
  \]

  \item \textbf{Soundness of Edge Application:} Each edge \( e = (v_j, v_{j+1}) \in T_i \) must correspond to a valid rule instance \( r \in R \) such that:
  \[
  r(K(v_j)) \vdash K(v_{j+1})
  \]
  Verifiability implies that the agent produces or references an explicit proof certificate \( \Pi_{j \rightarrow j+1} \) justifying the inference step.

  \item \textbf{Terminal Commitment:} The output \( \phi_i \) must be cryptographically committed and signed. If the agent holds private key \( sk_A \), then:
  \[
  \sigma_i = \text{Sign}_{sk_A}(H(\phi_i \| T_i \| \tau_i))
  \]
  with timestamp \( \tau_i \) and derivation trace \( T_i \) embedded in the signature context. Any verifying entity uses:
  \[
  \text{Verify}_{pk_A}(\sigma_i, H(\phi_i \| T_i \| \tau_i)) = \text{true}
  \]
\end{enumerate}

Collectively, these constraints define a formally verifiable output channel for reasoning agents. The enforcement mechanism assumes a verifier with access to the canonical Merkle root \( \text{Root}_G \), the set of inference rules \( R \), and the agent’s public key \( pk_A \). The entire validation protocol is executable on-chain if necessary, or via zero-knowledge circuits for privacy-preserving attestation.

Moreover, a failure in any verification step renders the output epistemically invalid—such outputs are considered hallucinations and must be disregarded in downstream reasoning. Thus, the DAG becomes not just a memory structure but a theorem-proving boundary for any acceptable cognitive operation in synthetic agents.

\section{Unforgeable Provenance Chains and Key Rotation Protocols}

To establish enduring trust in memory assertions and agent behaviour, each fragment of knowledge must be anchored in a verifiable provenance chain. In distributed epistemic systems, where actors evolve cryptographic identities over time, maintaining unforgeable provenance under dynamic key material necessitates formalised key rotation protocols with provable linkage.

Let each agent or user \( U \) be initially identified by a long-term public key \( pk_U^{(0)} \). As key pairs are rotated, we define a sequence of public keys \( \{pk_U^{(0)}, pk_U^{(1)}, \ldots, pk_U^{(n)}\} \), where each transition is cryptographically chained to its predecessor. Let the rotation certificate at epoch \( i \) be:

\[
C_U^{(i)} = \text{Sign}_{sk_U^{(i)}}(H(pk_U^{(i-1)} \parallel pk_U^{(i)} \parallel t_i))
\]

where \( t_i \) is a timestamp, and the hash binds the transition. The chain of such certificates \( \{C_U^{(1)}, \dots, C_U^{(n)}\} \) allows any verifier to recursively validate the evolution of the identity key. Verification is achieved by:

\[
\forall i \in [1,n], \quad \text{Verify}_{pk_U^{(i)}}(C_U^{(i)}, H(pk_U^{(i-1)} \parallel pk_U^{(i)} \parallel t_i)) = \text{true}
\]

This yields an unbroken, signed trail from \( pk_U^{(0)} \) to \( pk_U^{(n)} \), preserving the continuity of epistemic authorship.

For each knowledge fragment \( K_i \) contributed by agent \( U \), the commitment includes the signing key epoch \( e_i \) and digital signature:

\[
K_i = \{d_i, p_i, \sigma_i, e_i\}, \quad \sigma_i = \text{Sign}_{sk_U^{(e_i)}}(H(d_i \parallel p_i))
\]

To validate the provenance of \( K_i \), the verifier checks:

\begin{enumerate}
  \item The signature \( \sigma_i \) against \( pk_U^{(e_i)} \)
  \item The key rotation path from \( pk_U^{(0)} \) to \( pk_U^{(e_i)} \)
  \item The Merkle inclusion proof of \( H(d_i \parallel p_i) \) in the anchored ledger
\end{enumerate}

Thus, even after many rotations, the root of authorship is traceable to an initial cryptographic identity.

\subsection*{Forward-Secure Rotation and Revocation}

To prevent misuse of compromised keys, each epoch \( i \) includes metadata defining its deprecation horizon \( \delta_i \), a temporal or transaction-based bound beyond which the key is no longer trusted. Formally:

\[
\text{Valid}(pk_U^{(i)}) \iff t_{\text{use}} \in [t_i, t_i + \delta_i]
\]

Expired keys are pruned from verification contexts unless historical validation is required. Additionally, revocation lists or on-chain CRLs (Certificate Revocation Ledgers) provide tamper-evident publication of key withdrawal.

\subsection*{Merkle-Chained Rotation Ledger}

All rotation certificates \( C_U^{(i)} \) are committed in a Merkle tree \( M_U \), with root \( \mathcal{R}_U \) embedded in a public blockchain transaction. Verification of any key transition requires:

\begin{itemize}
  \item Inclusion proof \( \pi_i \) for \( C_U^{(i)} \) in \( M_U \)
  \item Verification that \( \mathcal{R}_U \) is anchored in a known block \( B \)
\end{itemize}

This architecture guarantees that key rotation cannot be forged or omitted, as absence invalidates the provenance path.

\subsection*{Immutability under Rotation}

Critically, the rotation of keys does not alter prior signed knowledge. Each fragment \( K_i \) remains permanently bound to the epoch key \( pk_U^{(e_i)} \), ensuring that agents cannot repudiate or retroactively alter past contributions even under new key pairs. The total structure thus achieves:

\begin{enumerate}
  \item Immutable authorship of memory fragments
  \item Forward security through key rotation
  \item Auditability via anchored Merkle proofs
  \item Revocability under formal expiration or compromise
\end{enumerate}

Such provenance chains, backed by cryptographic commitments and verifiable transition records, enable distributed agents to maintain continuity of trust without centralised authority, securing epistemic lineage even under adversarial threat conditions.

\section{Conclusion}

The architecture presented throughout this paper establishes a rigorous foundation for designing synthetic agents that are incapable of epistemic drift, hallucination, or unauthorised revision. By grounding memory and reasoning within cryptographic structures—specifically Merkle-rooted automata, authenticated knowledge graphs, and DAG-constrained traversal logic—agents transition from statistical artefacts to verifiable epistemic machines. Each assertion, memory fragment, and policy application is bound to an immutable historical context, externally auditable and cryptographically enforced.

Symmetric encryption derived from ECDH key exchanges ensures that access to information is not only authenticated but bound to hierarchical privilege structures. Append-only data semantics enforced through blockchain anchoring eliminates the possibility of destructive updates, and zero-knowledge proof systems allow selective, privacy-preserving audit of memory contents without data exposure. The enforcement of formal reasoning systems over a committed substrate guarantees that no output can be fabricated or detached from its logical derivation.

This reimagining of artificial cognition discards the legacy of probabilistic inference models that conflate approximation with understanding. Instead, it offers a cryptographically secure and ontologically grounded framework for agents whose knowledge is constrained by lawlike structures. Such agents do not think in the traditional sense; they reason under constraint. They cannot lie, cannot forget, and cannot feign belief. In this lies not mimicry of intelligence, but the foundation for institutional trust in synthetic cognition—an epistemology enforced by protocol rather than perception.

\newpage


\newpage


\begin{appendices}

\section{Key Derivation Functions: Formal Definitions}

A Key Derivation Function (KDF) is a deterministic algorithm that derives one or more secret keys from a source of initial keying material. In cryptographic systems designed for epistemic traceability and hierarchical access, KDFs serve as the mechanism by which contextual, privilege-bound, and time-anchored keys are generated. These derived keys ensure that memory fragments, agent assertions, and access controls remain cryptographically isolated and bound to their operational parameters.

Formally, a KDF is a function
\[
\text{KDF} : \{0,1\}^* \times \{0,1\}^* \rightarrow \{0,1\}^k
\]
mapping a pair of bitstrings—initial key material (IKM) and context string (CTX)—to an output of fixed length \( k \). A cryptographically secure KDF must exhibit pseudo-randomness, key separation, and domain separation, such that knowledge of one output does not yield information about any other.

We define the KDF function \( K_{\text{sym},i}^{(l)} \) used in our framework as:
\[
K_{\text{sym},i}^{(l)} = \text{HKDF}(K_{U,A} \parallel \text{context}_i \parallel l)
\]
where:

\begin{itemize}
    \item \( K_{U,A} \) is the shared secret derived via ECDH between the agent and a user,
    \item \( \text{context}_i \) encodes the semantic and temporal domain of the memory fragment,
    \item \( l \) is the privilege level from the access lattice \( L = \{l_1, ..., l_n\} \),
    \item \( \text{HKDF} \) is the HMAC-based KDF defined in RFC 5869.
\end{itemize}

HKDF operates as:
\[
\text{HKDF}(\text{IKM}, \text{salt}, \text{info}) = \text{OKM}
\]
where:

\begin{enumerate}
    \item The extraction phase:
    \[
    \text{PRK} = \text{HMAC}(\text{salt}, \text{IKM})
    \]
    \item The expansion phase:
    \[
    \text{OKM} = \text{Concat}(T_1, T_2, \dots, T_N)
    \]
    with
    \[
    T_1 = \text{HMAC}(\text{PRK}, \text{info} \parallel 0x01), \quad T_2 = \text{HMAC}(\text{PRK}, T_1 \parallel \text{info} \parallel 0x02), \dots
    \]
\end{enumerate}

The context string ensures derivational independence between key invocations. Even with a fixed ECDH shared secret, the output keys differ if their respective contexts differ, enforcing functional and ontological separation.

Moreover, by encoding time, role, and privilege directly within \( \text{context}_i \), the system guarantees that:
\begin{itemize}
    \item \textbf{Temporal integrity} is achieved: keys can be invalidated or scoped to specific time windows.
    \item \textbf{Role-based separation} ensures that distinct agents cannot share decryption ability even with the same shared secret.
    \item \textbf{Contextual immutability} binds the key to the semantic identity of the fragment—altering the context nullifies access.
\end{itemize}

In advanced implementations, HKDF may be replaced by quantum-resilient alternatives (e.g., SPHINCS+-based derivation mechanisms), with identical structure but post-quantum secure primitives. Nevertheless, the model of derivational unforgeability and semantic isolation remains invariant.

Thus, the KDF functions employed are not merely supportive cryptographic utilities; they are foundational mechanisms for epistemic confinement, access control, and identity differentiation within a provable cognitive substrate.

\section{ECDH Exchange Details}

Elliptic Curve Diffie-Hellman (ECDH) key exchange is the foundational mechanism through which agents and users establish a shared symmetric secret without direct transmission of the secret itself. The strength of ECDH lies in the computational hardness of the Elliptic Curve Discrete Logarithm Problem (ECDLP), which ensures that, even with full knowledge of the public keys, an adversary cannot feasibly derive the shared secret.

Let \( \mathbb{G} \) be a cyclic subgroup of an elliptic curve \( E \) over a finite field \( \mathbb{F}_q \), with generator point \( G \in E(\mathbb{F}_q) \) of prime order \( n \). Every participant possesses a private scalar and a corresponding public point.

Let:
\begin{itemize}
  \item User \( U \) have private key \( sk_U \in \mathbb{Z}_n \), and public key \( pk_U = sk_U \cdot G \),
  \item Agent \( A \) have private key \( sk_A \in \mathbb{Z}_n \), and public key \( pk_A = sk_A \cdot G \).
\end{itemize}

The shared secret is computed as:
\[
K_{U,A} = sk_U \cdot pk_A = sk_U \cdot (sk_A \cdot G) = sk_A \cdot pk_U
\]
This point \( K_{U,A} \in E(\mathbb{F}_q) \) lies on the curve and serves as a high-entropy input to a key derivation function (KDF), such as HKDF, which extracts a fixed-length symmetric key:
\[
K_{\text{shared}} = \text{HKDF}(x(K_{U,A}) \parallel \text{context})
\]
where \( x(K_{U,A}) \) denotes the \( x \)-coordinate of the elliptic curve point, and \( \text{context} \) incorporates protocol metadata (timestamps, privilege levels, Merkle roots).

This procedure ensures:
\begin{itemize}
  \item \textbf{Forward Secrecy:} Key material cannot be retroactively recovered if a long-term key is compromised.
  \item \textbf{Asymmetric Symmetry:} Both parties independently compute the same secret without sharing private keys.
  \item \textbf{Domain Separation:} The inclusion of \( \text{context} \) prevents key reuse across semantic domains.
\end{itemize}

Security assumptions rest on the infeasibility of solving:
\[
\text{Given } G, \quad A = a \cdot G, \quad B = b \cdot G, \quad \text{find } ab \cdot G.
\]
Without access to \( sk_U \) or \( sk_A \), the attacker faces the computational ECDLP, for which no polynomial-time classical algorithm exists.

The protocol is further hardened against man-in-the-middle attacks by cryptographically binding \( pk_U \) and \( pk_A \) to on-chain identities or digital signatures, optionally integrating zero-knowledge proofs to establish liveness and authorship without compromising secrecy.

In our framework, each fragment \( f_i \) is encrypted using a symmetric key derived from this shared secret:
\[
f_i = \{ \text{Enc}_{K_{\text{sym},i}}(d_i), H(d_i), \text{Proof}_{M_n}(H(d_i)) \}
\]
where \( K_{\text{sym},i} = \text{HKDF}(K_{U,A} \parallel \text{context}_i) \). Thus, ECDH ensures that only participants with valid elliptic curve credentials and aligned contextual privileges can decrypt or verify the fragment.

This exchange, paired with Merkle-rooted commitments, ensures data privacy, integrity, and provenance under cryptographic assumptions consistent with modern post-quantum migration pathways.

\section{Merkle Tree Inclusion Proofs}

Merkle trees provide an efficient and cryptographically secure method to verify the inclusion of a data element within a committed dataset, without revealing or transmitting the entire dataset. Let \( M_n \) denote the Merkle tree rooted in a known digest \( R_n \in \{0,1\}^\lambda \), where \( \lambda \) is the output length of the underlying cryptographic hash function \( H \). The tree is a full binary tree over a sequence of leaf values \( \{d_1, d_2, \ldots, d_{2^k} \} \), each hashed as \( h_i = H(d_i) \), with internal nodes recursively defined as \( H(h_L \parallel h_R) \) for children \( h_L \), \( h_R \).

Given a leaf \( d_i \), an inclusion proof \( \pi_i \) is defined as a sequence of sibling hashes on the path from \( h_i \) to the root:
\[
\pi_i = \left[ h^{(0)}_{s(1)}, h^{(1)}_{s(2)}, \ldots, h^{(k-1)}_{s(k)} \right]
\]
where \( s(j) \in \{ \text{left}, \text{right} \} \) specifies the sibling direction at level \( j \), and each \( h^{(j)} \) corresponds to a hash value at that level.

Verification is achieved via iterative reconstruction:
\[
h^{(0)} = H(d_i), \quad
h^{(j+1)} = \begin{cases}
H(h^{(j)} \parallel h^{(j)}_{s(j+1)}) & \text{if } s(j+1) = \text{right} \\
H(h^{(j)}_{s(j+1)} \parallel h^{(j)}) & \text{if } s(j+1) = \text{left}
\end{cases}
\]
until the final hash \( h^{(k)} \) is compared against \( R_n \). Inclusion is proven if and only if:
\[
h^{(k)} = R_n
\]

This mechanism provides:
\begin{itemize}
  \item \textbf{Logarithmic Efficiency:} Proofs are \( O(\log n) \) in size and computational cost.
  \item \textbf{Non-Interactive Verifiability:} The proof and the root are sufficient; the tree itself is not required.
  \item \textbf{Collision Resistance:} Security reduces to the preimage and collision resistance of \( H \), e.g., SHA-256.
\end{itemize}

In the proposed framework, each memory fragment \( f_i \) includes its own inclusion proof:
\[
f_i = \left\{ \text{Enc}_{K_{\text{sym},i}}(d_i), H(d_i), \pi_i \right\}
\]
which can be verified against the blockchain-anchored root \( R_n = \text{Root}(M_n) \) embedded in transaction \( T_n \) within block \( B_n \).

Thus, Merkle inclusion proofs enable decentralised agents and third parties to:
\begin{enumerate}
  \item Validate fragment membership in a committed knowledge set.
  \item Ensure immutability of agent memory without total disclosure.
  \item Support zero-knowledge protocols where the proof of inclusion suffices without revealing the fragment content.
\end{enumerate}

The epistemic implication is critical: no datum may be invoked, referenced, or reasoned upon unless its inclusion within a certified state is demonstrable. This prohibits hallucinations, enforces historical fidelity, and renders every memory access cryptographically traceable.

\section{Sample Blockchain Anchoring Schemas}

To ensure immutability and verifiability of agent memory, each critical state component—such as memory fragments, policy documents, or learned representations—is cryptographically committed to a blockchain ledger. The anchoring schema defines how these commitments are constructed, encoded, and published within blockchain transactions. The schema must ensure minimal bandwidth consumption, deterministic reconstruction, and robust tamper-evidence.

Let \( M_n \) denote a Merkle tree constructed at epoch \( n \), rooted in digest \( R_n = \text{Root}(M_n) \in \{0,1\}^\lambda \), where \( \lambda \) is the hash output length. The commitment schema \( \mathcal{A}_n \) for this epoch is embedded in a blockchain transaction \( T_n \), with the following data fields:

\begin{itemize}
  \item \textbf{Commitment Tag:} A structured identifier, e.g., \texttt{0xA1A1}, marking the transaction as a memory anchor.
  \item \textbf{Epoch ID:} A timestamp or logical sequence number \( t_n \in \mathbb{N} \), representing the anchoring round.
  \item \textbf{Merkle Root:} \( R_n \in \{0,1\}^\lambda \), committed as a fixed-length field in the transaction output.
  \item \textbf{Signature:} \( \sigma_n = \text{Sign}_{sk_A}(R_n \parallel t_n) \), signed by the agent’s private key \( sk_A \) for integrity and non-repudiation.
  \item \textbf{Optional Metadata:} Includes compression schemes, DAG indices, or access tags, optionally encrypted.
\end{itemize}

Thus, the full anchoring schema \( \mathcal{A}_n \) is:

\[
\mathcal{A}_n = \left( \texttt{tag}, t_n, R_n, \sigma_n, \mu_n \right)
\]

where \( \mu_n \) denotes optional metadata and auxiliary data.

\subsection*{Encoding in Blockchain Transactions}

For blockchains such as Bitcoin (BSV), the schema may be embedded in the \texttt{OP\_RETURN} field as:

\[
\texttt{OP\_RETURN} \ \texttt{tag} \parallel t_n \parallel R_n \parallel \sigma_n \parallel \mu_n
\]

This creates a publicly queryable and immutable commitment to \( M_n \) while preserving consensus validity.

\subsection*{Verification Protocol}

To validate the anchor, a verifier retrieves \( T_n \) from block \( B_n \), extracts \( \mathcal{A}_n \), and performs:

\begin{enumerate}
  \item Verify \( \sigma_n \) using agent’s public key \( pk_A \).
  \item Recompute Merkle root \( \hat{R}_n \) from fragment proofs \( \pi_i \).
  \item Compare \( \hat{R}_n = R_n \).
\end{enumerate}

If all steps succeed, inclusion and integrity are confirmed.

\subsection*{Schema Variants}

Anchoring schemas may vary based on operational constraints:

\begin{itemize}
  \item \textbf{Minimal Anchor:} Only the Merkle root \( R_n \) and epoch ID \( t_n \) are committed.
  \item \textbf{Signed Anchor:} Adds a digital signature \( \sigma_n \) for accountability.
  \item \textbf{Encrypted Anchor:} Metadata \( \mu_n \) is encrypted for access-tier encoding.
  \item \textbf{Batch Anchors:} Multiple roots \( \{R_n^1, R_n^2, \ldots\} \) included in a super-Merkle structure.
\end{itemize}

Each schema balances performance, legal verifiability, and cryptographic assurance. Anchoring transforms memory from mutable storage to a cryptographically witnessed journal of epistemic state.

\section{Access-Level Policy Graphs and Enforcement Maps}

In order to enforce structured access over fragmentary memory in synthetic agents, a formal system of access-level policy graphs is required. These graphs define which user identities, roles, or cryptographic keypairs may retrieve, decrypt, or traverse specific knowledge elements within the agent's memory graph. This design mandates a tiered enforcement mechanism grounded in lattice-based access control and mapped to cryptographic constructs.

\subsection*{Access Lattice Definition}

Let \( \mathcal{L} = \{ \ell_0, \ell_1, \ldots, \ell_m \} \) be a finite, totally ordered access lattice, such that:

\[
\ell_0 < \ell_1 < \ldots < \ell_m
\]

Each memory fragment \( f_i \) is associated with an access level \( \text{level}(f_i) \in \mathcal{L} \). Similarly, each user \( U_j \) possesses a clearance level \( \text{clearance}(U_j) \in \mathcal{L} \).

\subsection*{Policy Graph Structure}

Define the policy graph as a directed labelled graph:

\[
\mathcal{P} = (N, E)
\]

where:

\begin{itemize}
  \item \( N = \{ n_i \} \) is the set of memory fragments or logic nodes, each tagged with \( \text{level}(n_i) \in \mathcal{L} \).
  \item \( E = \{ (n_i, n_j, \rho_k) \} \) is a set of edges labelled by access predicates \( \rho_k \), representing traversal or transformation rights.
\end{itemize}

Traversal across edges is permitted only if the accessing user satisfies:

\[
\text{clearance}(U_j) \geq \max(\text{level}(n_i), \text{level}(n_j))
\]

This ensures no path may be navigated that violates the lattice ordering.

\subsection*{Enforcement Maps}

Enforcement is achieved via encryption-based segmentation of the graph. For each node \( n_i \), derive a symmetric key:

\[
K_i^{(\ell)} = \text{HKDF}(K_{U,A} \parallel \text{context}_i \parallel \ell)
\]

Only users with clearance \( \ell' \geq \ell \) and the appropriate shared key material \( K_{U,A} \) can derive \( K_i^{(\ell)} \) and decrypt the fragment \( f_i \). The node payload is:

\[
f_i = \left\{ \text{Enc}_{K_i^{(\ell)}}(d_i), H(d_i), \pi_i \right\}
\]

where \( \pi_i \) is the Merkle inclusion proof within block \( B_n \).

\subsection*{Graph Traversal and Inference Constraints}

Logical inference operations in the agent are constrained by clearance propagation. That is, if an agent’s internal reasoning seeks to traverse a path \( (n_a, n_b, n_c) \), then:

\[
\text{clearance}_A \geq \max \left( \text{level}(n_a), \text{level}(n_b), \text{level}(n_c) \right)
\]

Agents lacking such clearance are unable to execute or even symbolically model these inferences.

\subsection*{Security Properties}

This scheme yields several strong guarantees:

\begin{itemize}
  \item \textbf{Confidentiality:} Fragments are encrypted under level-specific keys.
  \item \textbf{Non-bypassability:} Graph traversal and output generation are bound by enforcement maps.
  \item \textbf{Auditability:} Each decryption and traversal is logged with cryptographic timestamping.
\end{itemize}

Access-level policy graphs enable robust, cryptographically enforced compartmentalisation of memory, allowing agents to process information selectively, in accordance with formal ontological privilege structures.

\end{appendices}

\end{document}